\documentclass[twocolumn,aps,showpacs,showkeys,bbm,prb,superscriptaddress]{revtex4-1}
\usepackage[latin9]{inputenc}
\usepackage{color}
\usepackage{units}
\usepackage{textcomp}
\usepackage{amsmath}
\usepackage{amssymb}
\usepackage{graphicx}
\usepackage[unicode=true,pdfusetitle,
 bookmarks=true,bookmarksnumbered=true,bookmarksopen=false,
 breaklinks=true,pdfborder={0 0 0},pdfborderstyle={},backref=false,colorlinks=true]
 {hyperref}
\hypersetup{
 pdfborderstyle={},pdfborderstyle={},pdfborderstyle={},pdfborderstyle={}}
\usepackage{breakurl}

\makeatletter
\usepackage{color}
\usepackage{bbm}
\usepackage{upgreek}

\AtBeginDocument{
  
}

\makeatother

\begin{document}

\title{High-efficiency optical pumping of nuclear polarization in a GaAs
quantum well}

\author{R. W. Mocek}
\email{raphael.mocek@tu-dortmund.de}
\affiliation{Experimental Physics 3, TU Dortmund, 44221 Dortmund, Germany}

\author{V. L. Korenev}
\affiliation{Ioffe Institute of the RAS, St. Petersburg, 194021, Russia}

\author{M. Bayer}
\affiliation{Ioffe Institute of the RAS, St. Petersburg, 194021, Russia}
\affiliation{Experimental Physics 2, TU Dortmund, 44221 Dortmund, Germany}

\author{M. Kotur}
\affiliation{Ioffe Institute of the RAS, St. Petersburg, 194021, Russia}

\author{R. I. Dzhioev}
\affiliation{Ioffe Institute of the RAS, St. Petersburg, 194021, Russia}

\author{D. O. Tolmachev}
\affiliation{Experimental Physics 3, TU Dortmund, 44221 Dortmund, Germany}

\author{G. Cascio}
\affiliation{Experimental Physics 3, TU Dortmund, 44221 Dortmund, Germany}

\author{K. V. Kavokin}
\affiliation{Ioffe Institute of the RAS, St. Petersburg, 194021, Russia}
\affiliation{Spin Optics Laboratory, St. Petersburg State University, 1, Ulianovskaya,
St. Petersburg, 198504, Russia}

\author{D. Suter}
\email{dieter.suter@tu-dortmund.de}
\affiliation{Experimental Physics 3, TU Dortmund, 44221 Dortmund, Germany}

\begin{abstract}
\noindent 
The dynamic polarization of nuclear spins by photoexcited electrons is studied in a high quality GaAs/AlGaAs quantum well. We find a surprisingly high efficiency of the spin transfer
from the electrons to the nuclei as reflected by a maximum nuclear field of 0.9~T in a tilted external magnetic field of 1~T strength only. This high efficiency is due to a low leakage of spin out
of the polarized nuclear system, because mechanisms of spin relaxation other than the hyperfine interaction are strongly suppressed, leading to a long nuclear relaxation time of up to 1000~s. 
A key ingredient to that end is the low impurity concentration inside the heterostructure, while the electrostatic potential from charged impurities in the surrounding barriers becomes screened through illumination by which
the spin relaxation time is increased compared to keeping the system
in the dark. This finding indicates a strategy for obtaining
high nuclear spin polarization as required for long-lasting carrier spin coherence. 
\end{abstract}



\maketitle
\textsl{Introduction.} The hyperfine interaction between conduction-band
electrons and lattice nuclei represents a major source of spin
decoherence, undesirable for applications in, for instance, spintronics. 
Unfortunately, in III-V semiconductors all nuclear species have non-zero spin momenta. 
A possibility of fighting the hyperfine interaction induced electron
spin decoherence is to reduce the nuclear spin fluctuations by imposing
a magnetic order on the nuclear spin system. One approach to reach this goal is nuclear self-polarization
imposing a spontaneous ordering via the electron-nuclear feedback\,\cite{d1972dynamic,korenev1999dynamic}. 
Another way is cooling of the nuclear spin system down to a few $\upmu$K to reach a phase transition into an ordered phase\,\cite{merkulov1982phase}. 
Further, dynamic polarization of nuclear spins via coupling with charge carriers\,\cite{khaetskii2002electron} counteracts also the spin decoherence. 
The dynamic polarization reduces the entropy of the nuclear spin system\,\cite{goldman1970spin}. Different
types of nuclear magnetic ordering can be achieved by lowering the
nuclear spin temperature using adiabatic demagnetization\,\cite{oja1997nuclear}
or nuclear magnetic resonance techniques\,\cite{goldman1970spin}. So far, nuclear ordering
has not been achieved in semiconductors because the required initial
polarization of the nuclear spins is high, about $70\thinspace\text{\%}$\,\cite{merkulov1982phase}.

Reaching that high nuclear spin polarization has proved to be a challenging
task. For example, in case of a $10\thinspace\text{nm}$ GaAs quantum
well (QW)  the authors\,\cite{Kalevich1990} reported an average nuclear
spin polarization of $\left<I\right>\approx0.07$, which corresponds to 
an effective nuclear magnetic field acting on the electron spins
(the Overhauser field) of $\approx0.25\thinspace\text{T}$. For dynamic
nuclear polarization in single quantum dots nuclear magnetic fields up to
$3\thinspace\text{T}$ were reported as result of the enhanced
hyperfine interaction in these structures\,\cite{PhysRevLett.98.026806,PhysRevB.74.245306,PhysRevB.76.201301,gammon2001electron}.
However, the nuclear polarization efficiency (the so-called leakage factor, see below) so far did not exceed 70\% in those studies.
Furthermore, the strain in (In,Ga)As quantum dots prevents the establishment
of equilibrium in the nuclear spin system\,\cite{maletinsky2009breakdown},
representing a major obstacle on the way to nuclear spin ordering.

In this work, we study the dynamic polarization of nuclear spins in
a wide, virtually unstrained GaAs quantum well in a tilted external magnetic field. The nuclear fields
that we achieve reach remarkably high strengths close to $0.9\thinspace\text{T}$.
This corresponds to a nuclear spin polarization efficiency above $90\thinspace\text{\%}$. The measured
dynamics of onset and decay of nuclear polarization indicate a strategy
towards high nuclear polarization approaching unity that may be applied also to other semiconductor structures.

\begin{figure}[h]
\noindent \centering{}\includegraphics[width=0.8\columnwidth]{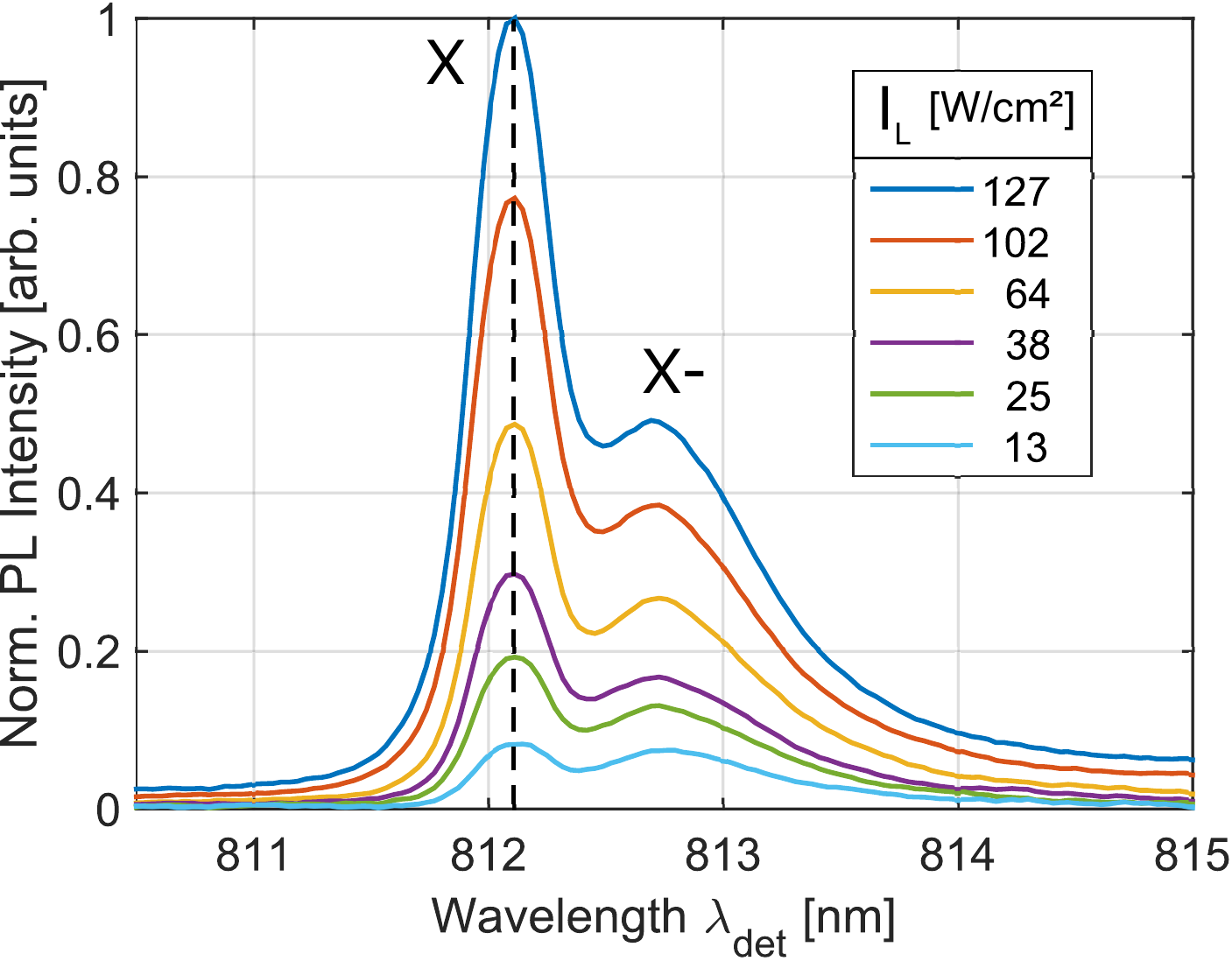}\caption{PL spectra of the $d=19.7\thinspace\text{nm}$ QW measured at temperature $T=1.6\text{K}$
for different excitation powers. The laser excitation wavelength was $\lambda_{\text{exc}}=800\,\text{nm}$. The vertical dashed line gives the detection
wavelength $\lambda_{\text{det}}$ for the subsequent measurements at the center of the e$_1$-hh$_1$ exciton (X).
\label{fig:experiment1}}
\end{figure}

\textsl{Sample.} The high quality sample used here was grown by molecular
beam epitaxy on a Te-doped GaAs substrate and consists of $13$ nominally undoped GaAs/Al$_{0.35}$Ga$_{0.65}$As QWs 
with thicknesses varing from $2.8$\,nm to $39.3$\,nm\,\cite{Eshlaghi08}. Figure\,\ref{fig:experiment1}
shows photoluminescence (PL) spectra in the energy range
where the emission of the $d=19.7\thinspace\text{nm}$ QW of interest occurs, measured at temperature $T=1.6\,\text{K}$ for different excitation intensities. 
The exciton emission line at $812.1$\,nm from the QW shows a width of $0.35$\,nm only, comparable to the high quality quantum wells studied in Ref.\,\cite{PhysRevB.84.155311}.
Its intensity increases considerably with excitation power, while the intensity of the additional feature at longer wavelengths remains almost constant as a decomposition of the signal into two spectral lines by a corresponding fit shows.  
We attribute it to emission from negatively charged excitons that are formed due to a small background doping in the barriers. 
However, we find no indication for emission from donor-bound excitons.

\begin{figure}
\noindent \begin{centering}
\includegraphics[width=0.8\columnwidth]{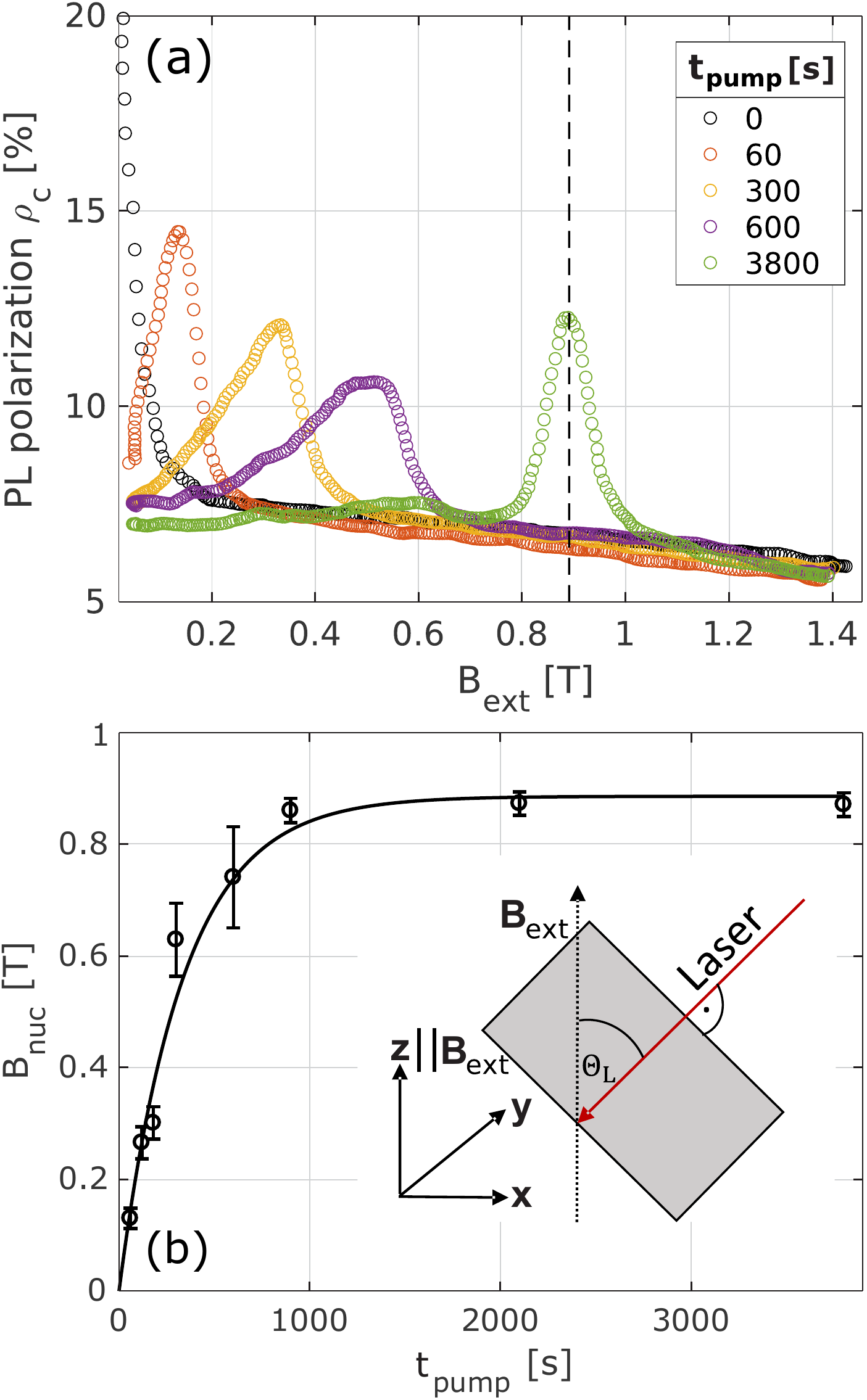}
\par\end{centering}
\caption{(a) Hanle curves measured after optically pumping the system for different
pumping times $t_{\text{pump}}$ at an external field $B_{\text{ext}}=1\thinspace\text{T}$,
a laser intensity $I_{\text{L}}=318\thinspace\text{W/cm}^2$, an optical
detuning $\Delta\lambda=0.35\thinspace\text{nm}$ and a temperature of $T=5.3\,\text{K}$. (b) Buildup of nuclear
magnetic field $B_{\text{nuc}}\left(t_{\text{pump}}\right)$ as function of 
pump time at $B_{\text{ext}}=1\thinspace\text{T}$. The symbols are
the Hanle curve maxima and the solid line is the fit using Eq.\,\eqref{eq:theory9}. The inset shows
the geometry of the experiment including the definition of the angle $\Theta_\text{L}$.
\label{fig:1THanleANDBnuc}}
\end{figure}

\textsl{Maximum achievable nuclear field.} First we measure the maximum achievable nuclear field in our structure. 
To that end, we summarize the Dyakonov-Perel formulation of dynamic nuclear
polarization (DNP)\,\cite{berkovits1973optical,DYAKONOV1984,d1974optical}. The kinetics of the average nuclear
spin are described by an exponential rise with the initial condition
$\left<I\left(t=0\right)\right>=0$: 
\begin{eqnarray}
\left<I\left(t\right)\right> & = & \left<I\right>_{\text{st}}\cdot\left(1-\exp\left(-\frac{t}{T_{1}}\right)\right),\label{eq:theory2}
\end{eqnarray}
where the nuclear relaxation time $T_{1}^{-1}=T_{\text{1e}}^{-1}+T_{\text{L}}^{-1}$
with $T_{\text{1e}}$ and $T_{\text{L}}$ being the relaxation times
via an electron and via other channels, e.g. quadrupole-induced relaxation, respectively. The
stationary value of the mean nuclear spin is given by 
\begin{eqnarray}
\left<I\right>_{\text{st}} & = & \frac{T_{\text{L}}}{T_{\text{1e}}+T_{\text{L}}}\cdot\frac{I\left(I+1\right)}{S\left(S+1\right)}S_{0}\cos\Theta_{\text{L}},\label{eq:theory5}
\end{eqnarray}
$I=\frac{3}{2}$ is the nuclear spin, $S=\frac{1}{2}$ is the electron
spin and $S_0$ is the electron spin polarization in absence of an external magnetic field. Here $\Theta_{\text{L}}$ is the angle between magnetic
field and optical axis, see  inset of Fig.\,\ref{fig:1THanleANDBnuc} (b). We use an oblique magnetic field to allow simultaneously
for DNP creation {\sl and} measurement of the Hanle effect in presence of an Overhauser field.

In addition, we introduce the ``leakage factor'' 
\begin{eqnarray}
f & = & \frac{T_{\text{L}}}{T_{\text{1e}}+T_{\text{L}}}.\label{eq:theory6}
\end{eqnarray}
If the spin-lattice relaxation is suppressed, $T_{\text{L}}$ will
exceed $T_{\text{1e}}$ by far ($T_{\text{L}}\gg T_{\text{1e}}$),
and the leakage factor approaches unity. The spin flow out of the
nuclear spin system is suppressed then, so that high nuclear polarization may be achieved. 
Vice versa, if the spin-lattice relaxation is efficient,
then $T_{\text{L}}\ll T_{\text{1e}}$ so that $f$ drops to zero.
Therefore the leakage factor is a good measure of the efficiency of
dynamic nuclear polarization by optically oriented electrons. Here,
we measure the kinetics of the buildup of the nuclear magnetic field
experimentally. The time dependence of the nuclear magnetic field
can be written as 
\begin{eqnarray}
B_{\text{nuc}}\left(t\right)=\frac{A\cdot\left<I\left(t\right)\right>}{\mu_{\text{B}}g_{\text{e}}}=B_{\text{nuc}}^{\text{st}}\cdot\left[1-\exp\left(-\frac{t}{T_{1}}\right)\right],\label{eq:theory9}
\end{eqnarray}
where $A$ is the hyperfine constant and $\mu_{\text{B}}$ is the Bohr magneton. The stationary value for the nuclear magnetic field is 
\begin{eqnarray}
B_{\text{nuc}}^{\text{st}}=B_{\text{nuc}}^{\text{max}}\cdot\frac{\left(I+1\right)}{S\left(S+1\right)}\cdot f\cdot S_{0}\cos\Theta_{\text{L}},\label{eq:theory11}
\end{eqnarray}
with $B_{\text{nuc}}^{\text{max}}=A\cdot I/\mu_{\text{B}}g_{\text{e}}=5.3\thinspace\text{T}$\,\cite{Paget1977}.
The electron g-factor in the $19.7\thinspace\text{nm}$ QW is close
to the one for bulk GaAs $g_{\text{e}}\approx0.4$\,\cite{Snelling1991}.
From the values for $T_{1}$ and $B_{\text{nuc}}^{\text{st}}$ extracted
from the experimental data using Eq.\,\eqref{eq:theory9}, we can
determine $T_{\text{L}}$ and $T_{\text{1e}}$.

\textsl{Experimental protocol.} We implement the optical orientation
method in a reflection geometry, where the sample is excited with
circularly polarized light to inject spin-polarized electrons\,\cite{Lampel1968,DYAKONOV1984,Abragam1961}.
We analyze the influence of the nuclear field on the circular PL polarization\,\cite{ekimov1970optical,Parsons1969,Dzhioev}
defined as $\rho_{\text{c}}=\left(I\left(\sigma_{+}\right)-I\left(\sigma_{-}\right)\right)/\left(I\left(\sigma_{+}\right)+I\left(\sigma_{-}\right)\right)$,
where $I\left(\sigma_{\pm}\right)$ is the intensity of the right-
or left-circularly polarized PL emission, respectively. The PL is
passed through a photo-elastic modulator and a monochromator
after which it is detected by an avalanche photodiode. The magnetic
field is generated by an electromagnet with field strengths up to
$B_{\text{ext}}=\unit[1.4]{T}$. The sample is mounted on the cold
finger of a flow-cryostat and kept at a temperature of $T\approx\unit[5.2\pm0.3]{K}$.
Two lock-in amplifiers are used to measure the sum and the difference
of left- and right-circularly polarized light, $I_{\Sigma/\Delta}=I\left(\sigma_{+}\right)\pm I\left(\sigma_{-}\right)$.
The optical pumping is done in an external magnetic field of $B_{\text{ext}}=\unit[1]{T}$.
A continuous wave (cw) diode laser is tuned to the excitation wavelength $\lambda_{\text{exc}}=811.75\thinspace\text{nm}$
using an excitation intensity of $I_{\text{L}}=318\thinspace\text{W/cm}^2$.
The laser spot size is about $100\,\upmu\text{m}$ for all measurements.
The monochromator is set to $\lambda_{\text{det}}=812.10\thinspace\text{nm}$,
which corresponds to the maximum of the e$_1$-hh$_1$ transition, see the
vertical dashed line in Fig.\,\ref{fig:experiment1}, and the optical
detuning of $\Delta\lambda=\lambda_{\text{det}}-\lambda_{\text{exc}}=0.35\thinspace\text{nm}$.

The measurement protocol consists of two steps. First we optically
pump the nuclear spin system and monitor the buildup of PL polarization.
Then we detect the nuclear spin polarization by measuring Hanle curves\,\cite{Hanle,DYAKONOV1984}
after different pumping times $t_{\text{pump}}$. To that end, the
magnetic field at $\Theta_{\text{L}}=73\pm2\text{\textdegree}$ is
scanned across the accessible field range from $0-1.4\thinspace\text{T}$.
The Hanle measurements are done at a reduced laser intensity of $I_{\text{L}}\approx\unit[24]{W/cm^2}$ and a scan-time of $\unit[25]{s}$, to minimize  within the technical limitations disturbances due to further optical pumping.

Figure\,\ref{fig:1THanleANDBnuc} (a) shows Hanle curves measured after pumping the system for different $t_{\text{pump}}$. The vertical
dashed line marks the center of the Hanle curve after the maximum
used pump time $t_{\text{pump}}=3800\thinspace\text{s}$, where we
observe a nuclear magnetic field of $B_{\text{nuc}}\left(t_{\text{pump}}=3800\thinspace\text{s}\right)=0.89\pm0.02\thinspace\text{T}$,
the maximum field achievable for the chosen conditions. All measurements were done at $T=5.3\,\text{K}$. The buildup
of the nuclear magnetic field $B_{\text{nuc}}\left(t_{\text{pump}}\right)$
is shown in Fig.\,\ref{fig:1THanleANDBnuc} (b) as function of pump
time. The solid line gives the fit according to Eq.~\eqref{eq:theory9}.
The fit parameters are: Nuclear relaxation time $T_{\text{1}}=335.3\pm26.0\thinspace\text{s}$,
nuclear magnetic field $B_{\text{nuc}}^{\text{st}}=0.89\pm0.03\thinspace\text{T}$,
nuclear spin-electron relaxation time $T_{\text{1e}}=391.1\pm32.4\thinspace\text{s}$,
nuclear spin-lattice relaxation time $T_{\text{L}}=2354.0\pm176.5\thinspace\text{s}$,
and leakage factor $f=0.86\pm0.09$. We highlight the nuclear magnetic
field of nearly $B_{\text{nuc}}^{\text{st}}\approx0.9\thinspace\text{T}$
for an applied external field of $B_{\text{ext}}=1\thinspace\text{T}$
as well as the leakage factor $f\approx0.9$ close to unity, indicating
highly efficient nuclear polarization. This high efficiency is reflected
by the nuclear spin-lattice time $T_{\text{L}}$, which is nearly
seven times longer than $T_{\text{1e}}$.

\textsl{Measurement of spin-lattice relaxation.} Further insight into
the low spin flow out of the nuclear spin bath can be taken from studies
in which the nuclear spin relaxation time $T_{\text{1,dark}}$ is measured
directly. To that end, the experimental protocol is slightly adapted.
The external magnetic field is varied from 0
to $105\thinspace\text{mT}$, which we can measure with an accuracy of $0.1$\,mT. The temperature is set to $10\thinspace\text{K}$
or $20\thinspace\text{K}$. A cw Ti:sapphire laser is tuned to $\lambda_{\text{exc}}=811.44\thinspace\text{nm}$
resulting in $\Delta\lambda=0.66\thinspace\text{nm}$ for $T=10\thinspace\text{K}$
and $\Delta\lambda=0.71\thinspace\text{nm}$ for $T=20\thinspace\text{K}$.
The laser intensity is $I_{\text{L}}=127\thinspace\text{W/cm}^2$.

To measure the nuclear spin-lattice relaxation time without the influence of spin-polarized electrons acting on the nuclei through an effective field known as Knight field, 
we use a three-stage protocol where the optical
pumping, the relaxation and the detection of the remaining nuclear
spin polarization are separated in time \,\cite{kotur2014nuclear,kotur2016nuclear}.
The first stage addresses the optical pumping of the nuclear spin
system in oblique magnetic field ($\Theta_{\text{L}}\geq80\text{\textdegree}$): $B_{\text{ext,light}}=35\thinspace\text{mT}$
for $T=10\thinspace\text{K}$ and $B_{\text{ext,light}}=70\thinspace\text{mT}$
for $T=20\thinspace\text{K}$. Thereafter the excitation beam is switched
off and the magnetic field is set to the $B_{\text{ext,dark}}$ at which
we want to determine $T_{\text{1,dark}}$ as function of the dark period
duration $t_{\text{dark}}$. During this dark period the nuclear field
decreases by the factor $\sim\exp\left(-t_{\text{\text{dark }}}/T_{\text{1,dark}}\right)$.
Right after the dark period the remaining nuclear polarization is
measured through the PL polarization representing the
third stage, where the laser beam is applied again and the external
magnetic field $B_{\text{ext,light}}$ is restored.

From the PL polarization at the beginning of the third stage $\rho_{\text{dark}}\left(t_{\text{dark}}\right)$
and using Eq.\,\eqref{eq:ExpBnucTdark}\,\cite{kotur2014nuclear}
we obtain information about the dynamics of the nuclear magnetic field
in the dark 
\begin{eqnarray}
B_{\text{nuc}}\left(t_{\text{dark}}\right) & = & B_{1/2}\sqrt{\frac{\rho_{\text{0}}-\rho_{\text{dark}}}{\rho_{\text{dark}}}}-B_{\text{ext,light}},\label{eq:ExpBnucTdark}
\end{eqnarray}
where $\rho_{0}$ is the degree of PL polarization in the absence
of an external magnetic field and $B_{1/2}$ is the half width at half maximum of the pure
electronic Hanle curve ($\sigma_{+}/\sigma_{-}$ excitation at $26\thinspace\text{kHz}$
modulation frequency). Once knowing the dependence $B_{\text{nuc}}\left(t_{\text{dark}}\right)$
we determine the relaxation time $T_{\text{1,dark}}$ from a fit with an
exponential decay function.

In order to compare the measured $T_{\text{1,dark}}$ times in the absence
of pumping with the time constant of nuclear polarization buildup,
in additional measurements the sample is kept in the dark at $B_{\text{ext,dark}}=0$
for $t_{\text{dark}}=50\thinspace\text{s}$ in order to completely
cancel the Overhauser field. Thereafter the optical pumping is resumed
for different external magnetic fields $B_{\text{ext, light}}$ and 
the time evolution of PL polarization due to the DNP is measured.

\begin{figure}[h]
\begin{centering}
\includegraphics[width=0.8\columnwidth]{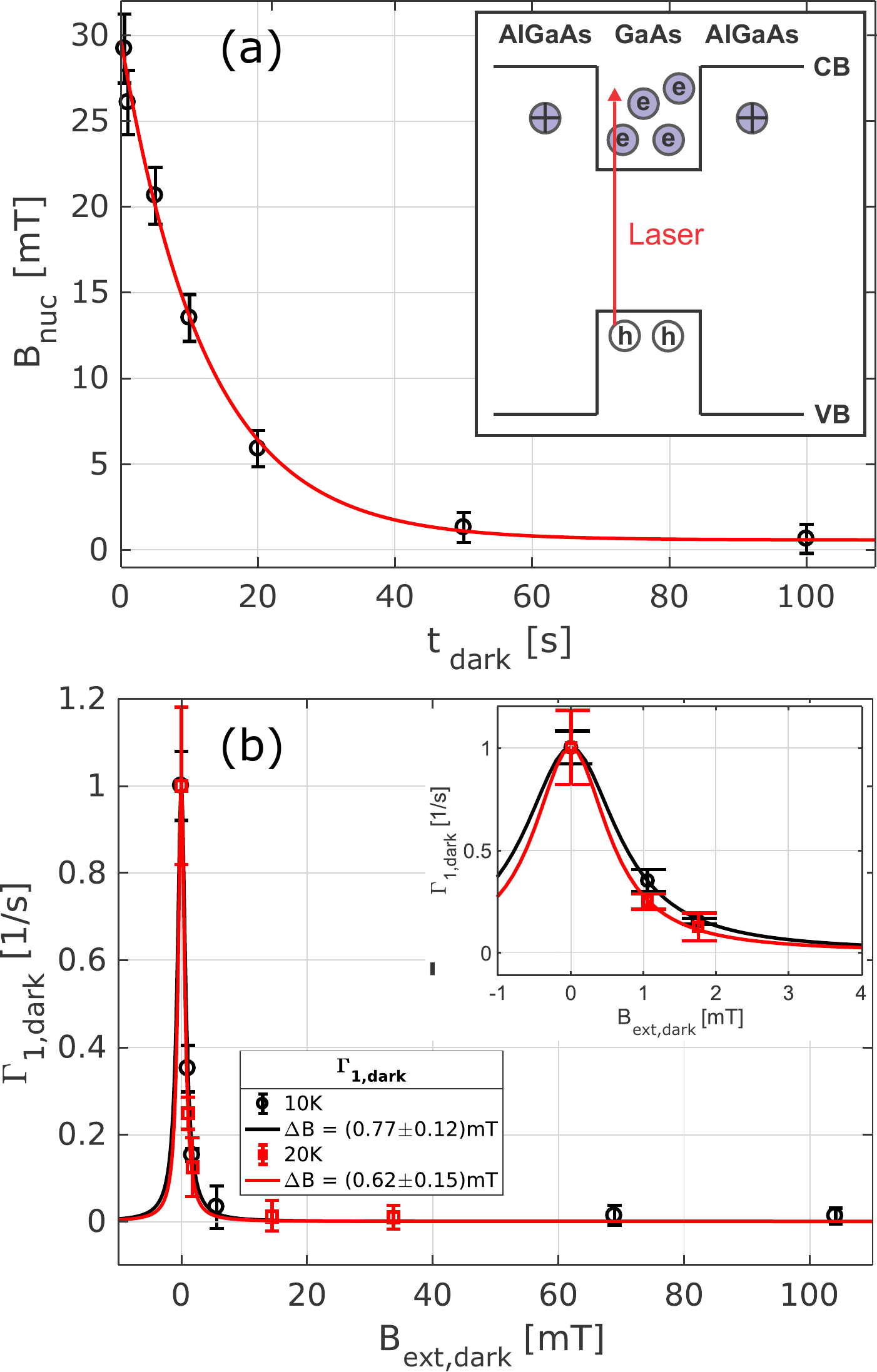}
\par\end{centering}
\caption{(a) Measured $B_{\text{nuc}}$ in dependence of the relaxation period
in darkness $t_{\text{dark}}$ at $B_{\text{ext,dark}}=1.8\pm0.2\thinspace\text{mT}$
and $T=10\thinspace\text{K}$. The solid red curve shows the result
of a fit using an exponential decay function with time constant $T_{\text{1,dark}}$. 
The inset shows schematically the origin of free electrons (e) in the GaAs QW by donors (grey pluses) 
mostly located in the AlGaAs barrier. 
Electrons and holes (h) are also excited by the laser in the conduction (CB) and valence (VB) bands.       
(b) Relaxation rates $\Gamma_{\text{1,dark}}$ \textsl{vs} magnetic
field in the dark for temperatures $T=10\thinspace\text{K},\thinspace20\thinspace\text{K}$.
Solid lines are Lorentzian fits to the experimental data. The inset
is a close-up of the data for $B_{\text{ext,dark}}=0-4\thinspace\text{mT}$.
\label{fig:BnucTdarkGammaDark}}
\end{figure}

\textsl{Results.} To understand the origin of the small leakage of
nuclear spin we compare the nuclear spin relaxation times $T_{1}$
both under pumping and in the dark. Figure\,\ref{fig:BnucTdarkGammaDark}
(a) shows the change of Overhauser field as function of
the dark relaxation period $t_{\text{dark}}$ at $T=10\thinspace\text{K}$
and $B_{\text{ext,dark}}=1.8\pm0.2\thinspace\text{mT}$. Fitting these
data with an exponential decay function gives us the $T_{\text{1,dark}}$,
which due to the absence of photoexcited carriers corresponds to the spin-lattice relaxation
time in darkness. We repeated this procedure for a set of dark relaxation
fields $B_{\text{ext,dark}}$ at $T=10\thinspace\text{K}$ and $T=20\thinspace\text{K}$.

\begin{figure}[h]
\begin{centering}
\includegraphics[width=0.8\columnwidth]{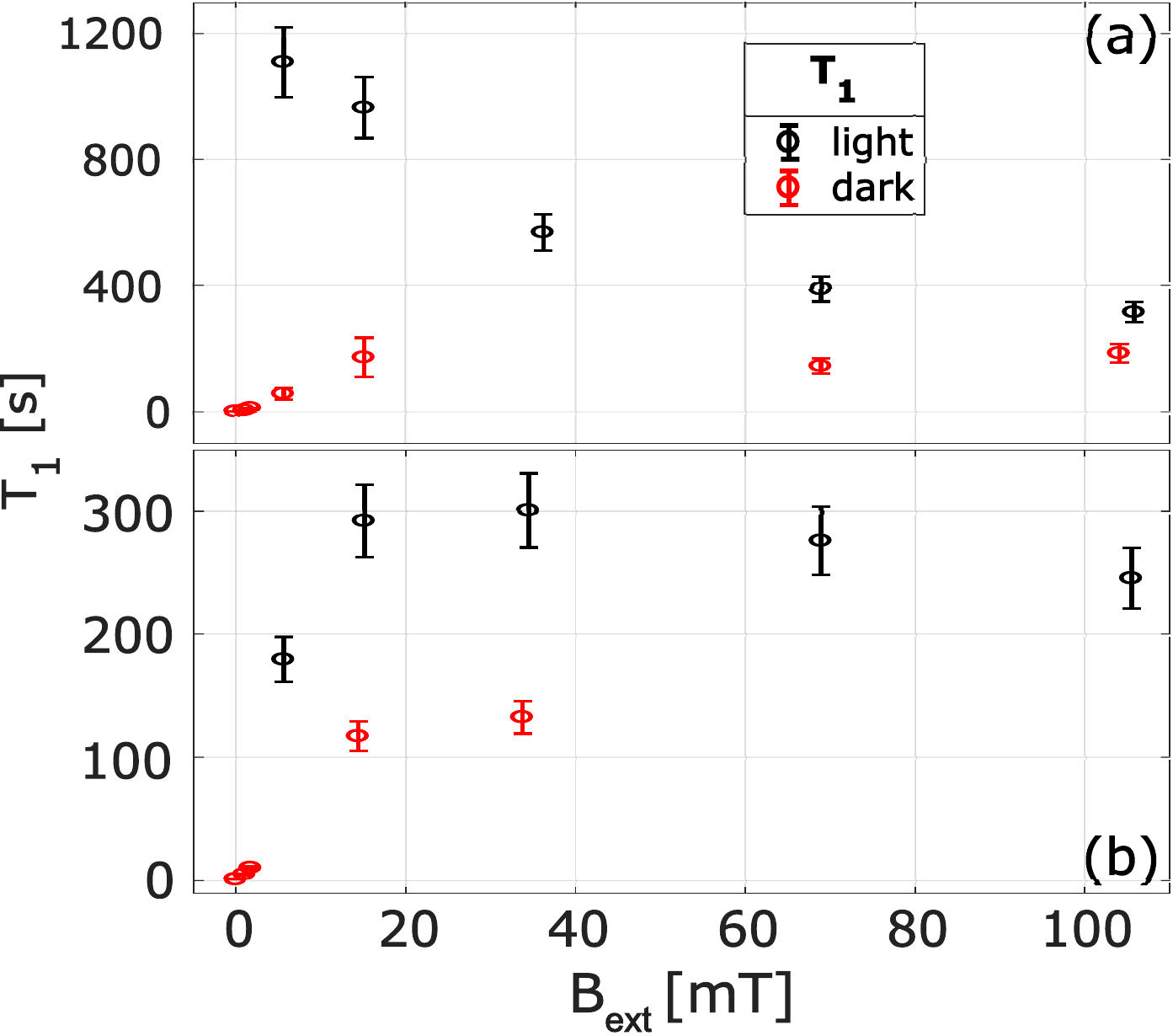} 
\par\end{centering}
\caption{Comparison of the dark and bright relaxation times $T_{\text{1,dark}}$
and $T_{\text{1,light}}$ measured at (a) $T=10\thinspace\text{K}$
and (b) $T=20\thinspace\text{K}$ {\sl vs} external magnetic field.
\label{fig:T1LightDark}}
\end{figure}

The complete set of measured relaxation rates $\Gamma_{\text{1,dark}}=T_{\text{1,dark}}^{-1}$
for different relaxation fields $B_{\text{ext,dark}}$ and the two temperatures
is shown in Fig.\,\ref{fig:BnucTdarkGammaDark} (b). The solid lines
are Lorentzian fits with a half width at half maximum of $\Delta B=0.8\pm0.1\thinspace\text{mT}$
for $T=10\thinspace\text{K}$ and $\Delta B=0.6\pm0.2\thinspace\text{mT}$
for $T=20\thinspace\text{K}$.
Figure\,\ref{fig:T1LightDark} shows the comparison of the dark and
bright (i.e. under illumination) relaxation times measured at (a)
$T=10\thinspace\text{K}$ and (b) $T=20\thinspace\text{K}$. Surprisingly,
we observe at least twice longer times for the nuclear spin polarization
buildup under illumination, which disturbs the system, than for the
spin polarization decay in darkness. 

We note that the observed trend for the nuclear relaxation times under illumination differs at $20$\,K in the low field regime from the one observed at $10$\,K. 
Namely, coming from high external fields one would expect a further increase of $T_{1,\text{light}}$ below $20$\,mT, while we observe a decrease. 
The source of this reduction is not fully understood yet and needs further investigation. A possible origin could be thermal activation of a further relaxation mechanism on these long time scales.
An example might be activation of weakly localized electrons in the quantum well plane, so that they become mobile. In higher magnetic fields, they may become re-localized by magnetic confinement.

\textsl{Discussion} The strong suppression of nuclear relaxation in
the dark already at weak magnetic fields of the order of $1\thinspace\text{mT}$
indicates a warm up of the nuclear spin system by slowly varying
electric fields via the quadrupole interaction for lower fields\,\cite{kotur2016nuclear}.
The characteristic field at which $T_{\text{1,dark}}$ increases twice
has to be interpreted then as the effective local field $B_{\text{L}}$
of the nuclear spin interactions. The extracted $B_{\text{L}}$ are a few times
larger than predicted for the dipole-dipole interaction between the
nuclear spins\,\cite{Paget1977}. They are typical for heterostructures and originate
from a static quadrupole splitting of the nuclear spin levels. The origin
may be some residual strain (as observed in Ref.\,\cite{PhysRevB.95.125312}
in a GaAs microcavity) or a static electric
field gradient.

For magnetic fields larger than $B_{\text{L}}$, the nuclear spin
relaxation rate is reduced by a factor of nearly $100$. Similar findings
were reported in Ref.\,\cite{kotur2016nuclear} for bulk n-GaAs,
however, with a reduction factor of about $10$ only. The reduction
was explained by suppression of the quadrupole relaxation, induced by slowly
fluctuating ($1$\,ms fluctuation time) donor charges\,\cite{kotur2014nuclear}. 
The low magnetic field required for suppression
indicates that there is only a small number of impurities in the QW,
as expected for our intrinsic, nominally undoped quantum well of high
quality, as evidenced by the narrow emission line. Otherwise such charges
would be the source of a background of field-independent relaxation
via hyperfine coupling. Therefore, the fluctuating electric fields
that induce the quadrupole warm up most likely originate from charges 
in the AlGaAs barriers\,\cite{doi:10.1063/1.355769}, see the sketch in Fig.\,\ref{fig:BnucTdarkGammaDark} (a).  For such impurities,
rechargement process were reported for excitation below the barriers\cite{PhysRevB.59.R10425,Volkov2000,PhysRevLett.88.256801}.

Having this in mind, one can explain the unusual finding that the
$T_{1}$-time is longer under pumping than in the dark. The long-range
fluctuating electric fields, which are created by charges in the AlGaAs barriers\,\cite{doi:10.1063/1.355769}, 
are screened by photoexcited carriers when the pump is on,
and the quadrupole warm-up is quenched. The contribution to nuclear
spin relaxation by residual localized electrons is negligible in our
i-type GaAs QW. The only remaining relaxation mechanism is hyperfine
scattering on free photoexcited two-dimensional electrons. The strength of this
mechanism was theoretically evaluated in Ref.\,\cite{Kalevich},
where the authors found that in case of non-degenerate free electrons
in a QW of width $d$ the hyperfine relaxation rate is 
\begin{equation}
T_{\text{1e}}^{\text{-1}}\propto \frac{A^{2}\Omega^{2}n_{\text{s}}m}{\hbar^{3}d^{2}},
\end{equation}
where $A$ is the hyperfine constant, $\Omega$ is the volume of the unit cell, $n_{\text{s}}$ is the QW electron concentration, $m$ is the effective mass of the electrons and $\hbar$ is the Dirac constant.  
For our $d=19.7\,\text{nm}$ QW this gives an estimate  in the order of $1000\thinspace\text{s}$, using $n_{\text{s}}=2\times10^{9}\,\text{cm}^{-2}$.
We highlight also the dependence of the relaxation time $T_{\text{1e}} \propto d^2$. 
This dependence is a consequence of the stronger electron confinement in narrower QWs, enhancing the hyperfine interaction with the nuclei and shortening the relaxation time. 
For this scaling to be valid, the QW width should however, not be too small, as the electrons then become localized in the QW plane.

Since the main mechanism of nuclear spin-lattice relaxation under
pumping turns out to be purely electronic, provided by the Fermi contact
interaction that conserves the total angular momentum of the interacting
particles, $T_{\text{1}}\approx T_{\text{1e}}$ leading to the low leakage of the nuclear spin 
as reflected by the value of $f$ close to unity\,\cite{Abragam1961}. 
This explains the strong Overhauser fields reached after long pumping in our experiments.

The presented results show that a wide undoped GaAs quantum well might
be the structure of choice for obtaining high degrees of nuclear spin
polarization by optical pumping. However, the fast quadrupole-induced
nuclear spin warm up observed at low magnetic fields in the dark may
hinder experiments targeting adiabatic demagnetization of the nuclear spins,
aimed at reaching ultralow spin temperatures and eventually nuclear
magnetic ordering. To reduce this undesirable effect, measures should
be taken to remove residual charged impurities outside the quantum well
and/or screening of fluctuating long-range electric fields in the
absence of optical pumping. This concept may be also transferred to
unstrained GaAs-based quantum dots that are appealing as host systems
for carrier spins suitable as quantum bits\,\cite{PMID:24500329,2014NatPh..10...46H}.

\textsl{Acknowledgments.} We acknowledge the financial support of
the Deutsche Forschungsgemeinschaft in the frame of the ICRC TRR 160
(Projects No. A6, C7) and the Russian Foundation for Basic Research
(Projects No. 15-52-12020, 15-52-12017).

 \bibliographystyle{apsrev}

\begin{thebibliography}{35}
\expandafter\ifx\csname natexlab\endcsname\relax\def\natexlab#1{#1}\fi
\expandafter\ifx\csname bibnamefont\endcsname\relax
  \def\bibnamefont#1{#1}\fi
\expandafter\ifx\csname bibfnamefont\endcsname\relax
  \def\bibfnamefont#1{#1}\fi
\expandafter\ifx\csname citenamefont\endcsname\relax
  \def\citenamefont#1{#1}\fi
\expandafter\ifx\csname url\endcsname\relax
  \def\url#1{\texttt{#1}}\fi
\expandafter\ifx\csname urlprefix\endcsname\relax\def\urlprefix{URL }\fi
\providecommand{\bibinfo}[2]{#2}
\providecommand{\eprint}[2][]{\url{#2}}

\bibitem[{\citenamefont{D'yakonov and Perel}(1972)}]{d1972dynamic}
\bibinfo{author}{\bibfnamefont{M.}~\bibnamefont{D'yakonov}} \bibnamefont{and}
  \bibinfo{author}{\bibfnamefont{V.}~\bibnamefont{Perel}},
  \bibinfo{journal}{JETP Lett.} \textbf{\bibinfo{volume}{16}},
  \bibinfo{pages}{398} (\bibinfo{year}{1972}).

\bibitem[{\citenamefont{Korenev}(1999)}]{korenev1999dynamic}
\bibinfo{author}{\bibfnamefont{V.}~\bibnamefont{Korenev}},
  \bibinfo{journal}{Zh. Eksp. Teor. Fiz.} \textbf{\bibinfo{volume}{70}},
  \bibinfo{pages}{129} (\bibinfo{year}{1999}).

\bibitem[{\citenamefont{Merkulov}(1982)}]{merkulov1982phase}
\bibinfo{author}{\bibfnamefont{I.}~\bibnamefont{Merkulov}},
  \bibinfo{journal}{Zh. Eksp. Teor. Fiz} \textbf{\bibinfo{volume}{82}},
  \bibinfo{pages}{319} (\bibinfo{year}{1982}).

\bibitem[{\citenamefont{Khaetskii et~al.}(2002)\citenamefont{Khaetskii, Loss,
  and Glazman}}]{khaetskii2002electron}
\bibinfo{author}{\bibfnamefont{A.~V.} \bibnamefont{Khaetskii}},
  \bibinfo{author}{\bibfnamefont{D.}~\bibnamefont{Loss}}, \bibnamefont{and}
  \bibinfo{author}{\bibfnamefont{L.}~\bibnamefont{Glazman}},
  \bibinfo{journal}{Phys. Rev. Lett.} \textbf{\bibinfo{volume}{88}},
  \bibinfo{pages}{186802} (\bibinfo{year}{2002}).

\bibitem[{\citenamefont{Goldman}(1970)}]{goldman1970spin}
\bibinfo{author}{\bibfnamefont{M.}~\bibnamefont{Goldman}},
  \emph{\bibinfo{title}{Spin temperature and nuclear magnetic resonance in
  solids}} (\bibinfo{publisher}{Clarendon Press Oxford}, \bibinfo{year}{1970}).

\bibitem[{\citenamefont{Oja and Lounasmaa}(1997)}]{oja1997nuclear}
\bibinfo{author}{\bibfnamefont{A.}~\bibnamefont{Oja}} \bibnamefont{and}
  \bibinfo{author}{\bibfnamefont{O.}~\bibnamefont{Lounasmaa}},
  \bibinfo{journal}{Rev. Mod. Phys.} \textbf{\bibinfo{volume}{69}},
  \bibinfo{pages}{1} (\bibinfo{year}{1997}).

\bibitem[{\citenamefont{Kalevich et~al.}(1990)\citenamefont{Kalevich, Korenev,
  and Fedorova}}]{Kalevich1990}
\bibinfo{author}{\bibfnamefont{V.}~\bibnamefont{Kalevich}},
  \bibinfo{author}{\bibfnamefont{V.}~\bibnamefont{Korenev}}, \bibnamefont{and}
  \bibinfo{author}{\bibfnamefont{O.}~\bibnamefont{Fedorova}},
  \bibinfo{journal}{JETP Lett.} \textbf{\bibinfo{volume}{52}},
  \bibinfo{pages}{349} (\bibinfo{year}{1990}).

\bibitem[{\citenamefont{Tartakovskii et~al.}(2007)\citenamefont{Tartakovskii,
  Wright, Russell, Fal'ko, Van'kov, Skiba-Szymanska, Drouzas, Kolodka,
  Skolnick, Fry et~al.}}]{PhysRevLett.98.026806}
\bibinfo{author}{\bibfnamefont{A.~I.} \bibnamefont{Tartakovskii}},
  \bibinfo{author}{\bibfnamefont{T.}~\bibnamefont{Wright}},
  \bibinfo{author}{\bibfnamefont{A.}~\bibnamefont{Russell}},
  \bibinfo{author}{\bibfnamefont{V.~I.} \bibnamefont{Fal'ko}},
  \bibinfo{author}{\bibfnamefont{A.~B.} \bibnamefont{Van'kov}},
  \bibinfo{author}{\bibfnamefont{J.}~\bibnamefont{Skiba-Szymanska}},
  \bibinfo{author}{\bibfnamefont{I.}~\bibnamefont{Drouzas}},
  \bibinfo{author}{\bibfnamefont{R.~S.} \bibnamefont{Kolodka}},
  \bibinfo{author}{\bibfnamefont{M.~S.} \bibnamefont{Skolnick}},
  \bibinfo{author}{\bibfnamefont{P.~W.} \bibnamefont{Fry}},
  \bibnamefont{et~al.}, \bibinfo{journal}{Phys. Rev. Lett.}
  \textbf{\bibinfo{volume}{98}}, \bibinfo{pages}{026806}
  (\bibinfo{year}{2007}).

\bibitem[{\citenamefont{Braun et~al.}(2006)\citenamefont{Braun, Urbaszek,
  Amand, Marie, Krebs, Eble, Lema{\^\i}tre, and Voisin}}]{PhysRevB.74.245306}
\bibinfo{author}{\bibfnamefont{P.-F.} \bibnamefont{Braun}},
  \bibinfo{author}{\bibfnamefont{B.}~\bibnamefont{Urbaszek}},
  \bibinfo{author}{\bibfnamefont{T.}~\bibnamefont{Amand}},
  \bibinfo{author}{\bibfnamefont{X.}~\bibnamefont{Marie}},
  \bibinfo{author}{\bibfnamefont{O.}~\bibnamefont{Krebs}},
  \bibinfo{author}{\bibfnamefont{B.}~\bibnamefont{Eble}},
  \bibinfo{author}{\bibfnamefont{A.}~\bibnamefont{Lema{\^\i}tre}},
  \bibnamefont{and} \bibinfo{author}{\bibfnamefont{P.}~\bibnamefont{Voisin}},
  \bibinfo{journal}{Phys. Rev. B} \textbf{\bibinfo{volume}{74}},
  \bibinfo{pages}{245306} (\bibinfo{year}{2006}).

\bibitem[{\citenamefont{Urbaszek et~al.}(2007)\citenamefont{Urbaszek, Braun,
  Amand, Krebs, Belhadj, Lema{\^\i}tre, Voisin, and
  Marie}}]{PhysRevB.76.201301}
\bibinfo{author}{\bibfnamefont{B.}~\bibnamefont{Urbaszek}},
  \bibinfo{author}{\bibfnamefont{P.-F.} \bibnamefont{Braun}},
  \bibinfo{author}{\bibfnamefont{T.}~\bibnamefont{Amand}},
  \bibinfo{author}{\bibfnamefont{O.}~\bibnamefont{Krebs}},
  \bibinfo{author}{\bibfnamefont{T.}~\bibnamefont{Belhadj}},
  \bibinfo{author}{\bibfnamefont{A.}~\bibnamefont{Lema{\^\i}tre}},
  \bibinfo{author}{\bibfnamefont{P.}~\bibnamefont{Voisin}}, \bibnamefont{and}
  \bibinfo{author}{\bibfnamefont{X.}~\bibnamefont{Marie}},
  \bibinfo{journal}{Phys. Rev. B} \textbf{\bibinfo{volume}{76}},
  \bibinfo{pages}{201301} (\bibinfo{year}{2007}).

\bibitem[{\citenamefont{Gammon et~al.}(2001)\citenamefont{Gammon, Efros,
  Kennedy, Rosen, Katzer, Park, Brown, Korenev, and
  Merkulov}}]{gammon2001electron}
\bibinfo{author}{\bibfnamefont{D.}~\bibnamefont{Gammon}},
  \bibinfo{author}{\bibfnamefont{A.~L.} \bibnamefont{Efros}},
  \bibinfo{author}{\bibfnamefont{T.}~\bibnamefont{Kennedy}},
  \bibinfo{author}{\bibfnamefont{M.}~\bibnamefont{Rosen}},
  \bibinfo{author}{\bibfnamefont{D.}~\bibnamefont{Katzer}},
  \bibinfo{author}{\bibfnamefont{D.}~\bibnamefont{Park}},
  \bibinfo{author}{\bibfnamefont{S.~W.} \bibnamefont{Brown}},
  \bibinfo{author}{\bibfnamefont{V.}~\bibnamefont{Korenev}}, \bibnamefont{and}
  \bibinfo{author}{\bibfnamefont{I.}~\bibnamefont{Merkulov}},
  \bibinfo{journal}{Phys. Rev. Lett.} \textbf{\bibinfo{volume}{86}},
  \bibinfo{pages}{5176} (\bibinfo{year}{2001}).

\bibitem[{\citenamefont{Maletinsky et~al.}(2009)\citenamefont{Maletinsky,
  Kroner, and Imamoglu}}]{maletinsky2009breakdown}
\bibinfo{author}{\bibfnamefont{P.}~\bibnamefont{Maletinsky}},
  \bibinfo{author}{\bibfnamefont{M.}~\bibnamefont{Kroner}}, \bibnamefont{and}
  \bibinfo{author}{\bibfnamefont{A.}~\bibnamefont{Imamoglu}},
  \bibinfo{journal}{Nat. Phys.} \textbf{\bibinfo{volume}{5}},
  \bibinfo{pages}{407} (\bibinfo{year}{2009}).

\bibitem[{\citenamefont{Eshlaghi et~al.}(2008)\citenamefont{Eshlaghi, Worthoff,
  Wieck, and Suter}}]{Eshlaghi08}
\bibinfo{author}{\bibfnamefont{S.}~\bibnamefont{Eshlaghi}},
  \bibinfo{author}{\bibfnamefont{W.}~\bibnamefont{Worthoff}},
  \bibinfo{author}{\bibfnamefont{A.~D.} \bibnamefont{Wieck}}, \bibnamefont{and}
  \bibinfo{author}{\bibfnamefont{D.}~\bibnamefont{Suter}},
  \bibinfo{journal}{Phys. Rev. B} \textbf{\bibinfo{volume}{77}},
  \bibinfo{pages}{245317} (\bibinfo{year}{2008}).

\bibitem[{\citenamefont{Moret et~al.}(2011)\citenamefont{Moret, Oberli,
  Pelucchi, Gogneau, Rudra, and Kapon}}]{PhysRevB.84.155311}
\bibinfo{author}{\bibfnamefont{N.}~\bibnamefont{Moret}},
  \bibinfo{author}{\bibfnamefont{D.~Y.} \bibnamefont{Oberli}},
  \bibinfo{author}{\bibfnamefont{E.}~\bibnamefont{Pelucchi}},
  \bibinfo{author}{\bibfnamefont{N.}~\bibnamefont{Gogneau}},
  \bibinfo{author}{\bibfnamefont{A.}~\bibnamefont{Rudra}}, \bibnamefont{and}
  \bibinfo{author}{\bibfnamefont{E.}~\bibnamefont{Kapon}},
  \bibinfo{journal}{Phys. Rev. B} \textbf{\bibinfo{volume}{84}},
  \bibinfo{pages}{155311} (\bibinfo{year}{2011}).

\bibitem[{\citenamefont{Dyakonov and Perel}(1974)}]{berkovits1973optical}
\bibinfo{author}{\bibfnamefont{M.}~\bibnamefont{Dyakonov}} \bibnamefont{and}
  \bibinfo{author}{\bibfnamefont{V.}~\bibnamefont{Perel}},
  \bibinfo{journal}{JETP} \textbf{\bibinfo{volume}{38}}, \bibinfo{pages}{177}
  (\bibinfo{year}{1974}).

\bibitem[{\citenamefont{Dyakonov and Perel}(1984)}]{DYAKONOV1984}
\bibinfo{author}{\bibfnamefont{M.}~\bibnamefont{Dyakonov}} \bibnamefont{and}
  \bibinfo{author}{\bibfnamefont{V.}~\bibnamefont{Perel}}, in
  \emph{\bibinfo{booktitle}{Optical Orientation}}, edited by
  \bibinfo{editor}{\bibfnamefont{F.}~\bibnamefont{Meier}} \bibnamefont{and}
  \bibinfo{editor}{\bibfnamefont{B.}~\bibnamefont{Zakharchenya}}
  (\bibinfo{publisher}{Elsevier}, \bibinfo{address}{Amsterdam},
  \bibinfo{year}{1984}), vol.~\bibinfo{volume}{8} of
  \emph{\bibinfo{series}{Modern Problems in Condensed Matter Sciences}}, pp.
  \bibinfo{pages}{11 -- 71}.

\bibitem[{\citenamefont{Dyakonov et~al.}(1975)\citenamefont{Dyakonov, Perel,
  Berkovits, and Safarov}}]{d1974optical}
\bibinfo{author}{\bibfnamefont{M.}~\bibnamefont{Dyakonov}},
  \bibinfo{author}{\bibfnamefont{V.}~\bibnamefont{Perel}},
  \bibinfo{author}{\bibfnamefont{V.}~\bibnamefont{Berkovits}},
  \bibnamefont{and} \bibinfo{author}{\bibfnamefont{V.}~\bibnamefont{Safarov}},
  \bibinfo{journal}{JETP} \textbf{\bibinfo{volume}{40}}, \bibinfo{pages}{950}
  (\bibinfo{year}{1975}).

\bibitem[{\citenamefont{Paget et~al.}(1977)\citenamefont{Paget, Lampel,
  Sapoval, and Safarov}}]{Paget1977}
\bibinfo{author}{\bibfnamefont{D.}~\bibnamefont{Paget}},
  \bibinfo{author}{\bibfnamefont{G.}~\bibnamefont{Lampel}},
  \bibinfo{author}{\bibfnamefont{B.}~\bibnamefont{Sapoval}}, \bibnamefont{and}
  \bibinfo{author}{\bibfnamefont{V.~I.} \bibnamefont{Safarov}},
  \bibinfo{journal}{Phys. Rev. B} \textbf{\bibinfo{volume}{15}},
  \bibinfo{pages}{5780} (\bibinfo{year}{1977}).

\bibitem[{\citenamefont{Snelling et~al.}(1991)\citenamefont{Snelling, Flinn,
  Plaut, Harley, Tropper, Eccleston, and Phillips}}]{Snelling1991}
\bibinfo{author}{\bibfnamefont{M.}~\bibnamefont{Snelling}},
  \bibinfo{author}{\bibfnamefont{G.}~\bibnamefont{Flinn}},
  \bibinfo{author}{\bibfnamefont{A.}~\bibnamefont{Plaut}},
  \bibinfo{author}{\bibfnamefont{R.}~\bibnamefont{Harley}},
  \bibinfo{author}{\bibfnamefont{A.}~\bibnamefont{Tropper}},
  \bibinfo{author}{\bibfnamefont{R.}~\bibnamefont{Eccleston}},
  \bibnamefont{and} \bibinfo{author}{\bibfnamefont{C.}~\bibnamefont{Phillips}},
  \bibinfo{journal}{Phys. Rev. B} \textbf{\bibinfo{volume}{44}},
  \bibinfo{pages}{11345} (\bibinfo{year}{1991}).

\bibitem[{\citenamefont{Lampel}(1968)}]{Lampel1968}
\bibinfo{author}{\bibfnamefont{G.}~\bibnamefont{Lampel}},
  \bibinfo{journal}{Phys. Rev. Lett.} \textbf{\bibinfo{volume}{20}},
  \bibinfo{pages}{491} (\bibinfo{year}{1968}).

\bibitem[{\citenamefont{Abragam}(1967)}]{Abragam1961}
\bibinfo{author}{\bibfnamefont{A.}~\bibnamefont{Abragam}},
  \emph{\bibinfo{title}{The Principles of Nuclear Magnetism}}
  (\bibinfo{publisher}{Clarendon Press}, \bibinfo{address}{Oxford},
  \bibinfo{year}{1967}).

\bibitem[{\citenamefont{Ekimov and Safarov}(1970)}]{ekimov1970optical}
\bibinfo{author}{\bibfnamefont{A.}~\bibnamefont{Ekimov}} \bibnamefont{and}
  \bibinfo{author}{\bibfnamefont{V.}~\bibnamefont{Safarov}},
  \bibinfo{journal}{JETP Lett.} \textbf{\bibinfo{volume}{12}},
  \bibinfo{pages}{1} (\bibinfo{year}{1970}).

\bibitem[{\citenamefont{Parsons}(1969)}]{Parsons1969}
\bibinfo{author}{\bibfnamefont{R.~R.} \bibnamefont{Parsons}},
  \bibinfo{journal}{Phys. Rev. Lett.} \textbf{\bibinfo{volume}{23}},
  \bibinfo{pages}{1152} (\bibinfo{year}{1969}).

\bibitem[{\citenamefont{Dzhioev et~al.}(1998)\citenamefont{Dzhioev,
  Zakharchenya, Korenev, Pak, Vinokurov, Kovalenkov, and Tarasov}}]{Dzhioev}
\bibinfo{author}{\bibfnamefont{R.~I.} \bibnamefont{Dzhioev}},
  \bibinfo{author}{\bibfnamefont{B.~P.} \bibnamefont{Zakharchenya}},
  \bibinfo{author}{\bibfnamefont{V.~L.} \bibnamefont{Korenev}},
  \bibinfo{author}{\bibfnamefont{P.~E.} \bibnamefont{Pak}},
  \bibinfo{author}{\bibfnamefont{D.~A.} \bibnamefont{Vinokurov}},
  \bibinfo{author}{\bibfnamefont{O.~V.} \bibnamefont{Kovalenkov}},
  \bibnamefont{and} \bibinfo{author}{\bibfnamefont{I.~S.}
  \bibnamefont{Tarasov}}, \bibinfo{journal}{Phys. Solid State}
  \textbf{\bibinfo{volume}{40}}, \bibinfo{pages}{1587} (\bibinfo{year}{1998}).

\bibitem[{\citenamefont{Hanle}(1924)}]{Hanle}
\bibinfo{author}{\bibfnamefont{W.}~\bibnamefont{Hanle}}, \bibinfo{journal}{Z.
  Phys.} \textbf{\bibinfo{volume}{30}}, \bibinfo{pages}{93}
  (\bibinfo{year}{1924}).

\bibitem[{\citenamefont{Kotur et~al.}(2014)\citenamefont{Kotur, Dzhioev,
  Kavokin, Korenev, Namozov, Pak, and Kusrayev}}]{kotur2014nuclear}
\bibinfo{author}{\bibfnamefont{M.}~\bibnamefont{Kotur}},
  \bibinfo{author}{\bibfnamefont{R.}~\bibnamefont{Dzhioev}},
  \bibinfo{author}{\bibfnamefont{K.}~\bibnamefont{Kavokin}},
  \bibinfo{author}{\bibfnamefont{V.~L.} \bibnamefont{Korenev}},
  \bibinfo{author}{\bibfnamefont{B.}~\bibnamefont{Namozov}},
  \bibinfo{author}{\bibfnamefont{P.}~\bibnamefont{Pak}}, \bibnamefont{and}
  \bibinfo{author}{\bibfnamefont{Y.~G.} \bibnamefont{Kusrayev}},
  \bibinfo{journal}{JETP Lett.} \textbf{\bibinfo{volume}{99}},
  \bibinfo{pages}{40} (\bibinfo{year}{2014}).

\bibitem[{\citenamefont{Kotur et~al.}(2016)\citenamefont{Kotur, Dzhioev,
  Vladimirova, Jouault, Korenev, and Kavokin}}]{kotur2016nuclear}
\bibinfo{author}{\bibfnamefont{M.}~\bibnamefont{Kotur}},
  \bibinfo{author}{\bibfnamefont{R.}~\bibnamefont{Dzhioev}},
  \bibinfo{author}{\bibfnamefont{M.}~\bibnamefont{Vladimirova}},
  \bibinfo{author}{\bibfnamefont{B.}~\bibnamefont{Jouault}},
  \bibinfo{author}{\bibfnamefont{V.}~\bibnamefont{Korenev}}, \bibnamefont{and}
  \bibinfo{author}{\bibfnamefont{K.}~\bibnamefont{Kavokin}},
  \bibinfo{journal}{Phys. Rev. B} \textbf{\bibinfo{volume}{94}},
  \bibinfo{pages}{081201} (\bibinfo{year}{2016}).

\bibitem[{\citenamefont{Vladimirova et~al.}(2017)\citenamefont{Vladimirova,
  Cronenberger, Scalbert, Kotur, Dzhioev, Ryzhov, Kozlov, Zapasskii,
  Lema{\^\i}tre, and Kavokin}}]{PhysRevB.95.125312}
\bibinfo{author}{\bibfnamefont{M.}~\bibnamefont{Vladimirova}},
  \bibinfo{author}{\bibfnamefont{S.}~\bibnamefont{Cronenberger}},
  \bibinfo{author}{\bibfnamefont{D.}~\bibnamefont{Scalbert}},
  \bibinfo{author}{\bibfnamefont{M.}~\bibnamefont{Kotur}},
  \bibinfo{author}{\bibfnamefont{R.~I.} \bibnamefont{Dzhioev}},
  \bibinfo{author}{\bibfnamefont{I.~I.} \bibnamefont{Ryzhov}},
  \bibinfo{author}{\bibfnamefont{G.~G.} \bibnamefont{Kozlov}},
  \bibinfo{author}{\bibfnamefont{V.~S.} \bibnamefont{Zapasskii}},
  \bibinfo{author}{\bibfnamefont{A.}~\bibnamefont{Lema{\^\i}tre}},
  \bibnamefont{and} \bibinfo{author}{\bibfnamefont{K.~V.}
  \bibnamefont{Kavokin}}, \bibinfo{journal}{Phys. Rev. B}
  \textbf{\bibinfo{volume}{95}}, \bibinfo{pages}{125312}
  (\bibinfo{year}{2017}).

\bibitem[{\citenamefont{Pavesi and Guzzi}(1994)}]{doi:10.1063/1.355769}
\bibinfo{author}{\bibfnamefont{L.}~\bibnamefont{Pavesi}} \bibnamefont{and}
  \bibinfo{author}{\bibfnamefont{M.}~\bibnamefont{Guzzi}}, \bibinfo{journal}{J.
  Appl. Phys.} \textbf{\bibinfo{volume}{75}}, \bibinfo{pages}{4779}
  (\bibinfo{year}{1994}).

\bibitem[{\citenamefont{Glasberg et~al.}(1999)\citenamefont{Glasberg,
  Finkelstein, Shtrikman, and Bar-Joseph}}]{PhysRevB.59.R10425}
\bibinfo{author}{\bibfnamefont{S.}~\bibnamefont{Glasberg}},
  \bibinfo{author}{\bibfnamefont{G.}~\bibnamefont{Finkelstein}},
  \bibinfo{author}{\bibfnamefont{H.}~\bibnamefont{Shtrikman}},
  \bibnamefont{and}
  \bibinfo{author}{\bibfnamefont{I.}~\bibnamefont{Bar-Joseph}},
  \bibinfo{journal}{Phys. Rev. B} \textbf{\bibinfo{volume}{59}},
  \bibinfo{pages}{R10425} (\bibinfo{year}{1999}).

\bibitem[{\citenamefont{Volkov et~al.}(2000)\citenamefont{Volkov, Kukushkin,
  Kulakovskii, von Klitzing, and Eberl}}]{Volkov2000}
\bibinfo{author}{\bibfnamefont{O.~V.} \bibnamefont{Volkov}},
  \bibinfo{author}{\bibfnamefont{I.~V.} \bibnamefont{Kukushkin}},
  \bibinfo{author}{\bibfnamefont{D.~V.} \bibnamefont{Kulakovskii}},
  \bibinfo{author}{\bibfnamefont{K.}~\bibnamefont{von Klitzing}},
  \bibnamefont{and} \bibinfo{author}{\bibfnamefont{K.}~\bibnamefont{Eberl}},
  \bibinfo{journal}{JETP Lett.} \textbf{\bibinfo{volume}{71}},
  \bibinfo{pages}{322} (\bibinfo{year}{2000}).

\bibitem[{\citenamefont{Dzhioev et~al.}(2002)\citenamefont{Dzhioev, Korenev,
  Merkulov, Zakharchenya, Gammon, Efros, and Katzer}}]{PhysRevLett.88.256801}
\bibinfo{author}{\bibfnamefont{R.~I.} \bibnamefont{Dzhioev}},
  \bibinfo{author}{\bibfnamefont{V.~L.} \bibnamefont{Korenev}},
  \bibinfo{author}{\bibfnamefont{I.~A.} \bibnamefont{Merkulov}},
  \bibinfo{author}{\bibfnamefont{B.~P.} \bibnamefont{Zakharchenya}},
  \bibinfo{author}{\bibfnamefont{D.}~\bibnamefont{Gammon}},
  \bibinfo{author}{\bibfnamefont{A.~L.} \bibnamefont{Efros}}, \bibnamefont{and}
  \bibinfo{author}{\bibfnamefont{D.~S.} \bibnamefont{Katzer}},
  \bibinfo{journal}{Phys. Rev. Lett.} \textbf{\bibinfo{volume}{88}},
  \bibinfo{pages}{256801} (\bibinfo{year}{2002}).

\bibitem[{\citenamefont{Kalevich and Korenev}(1991)}]{Kalevich}
\bibinfo{author}{\bibfnamefont{V.~K.} \bibnamefont{Kalevich}} \bibnamefont{and}
  \bibinfo{author}{\bibfnamefont{V.~L.} \bibnamefont{Korenev}},
  \bibinfo{journal}{Appl. Magn. Reson.} \textbf{\bibinfo{volume}{2}},
  \bibinfo{pages}{397} (\bibinfo{year}{1991}).

\bibitem[{\citenamefont{Sallen et~al.}(2014)\citenamefont{Sallen, Kunz, Amand,
  Bouet, Kuroda, Mano, Paget, Krebs, Marie, Sakoda et~al.}}]{PMID:24500329}
\bibinfo{author}{\bibfnamefont{G.}~\bibnamefont{Sallen}},
  \bibinfo{author}{\bibfnamefont{S.}~\bibnamefont{Kunz}},
  \bibinfo{author}{\bibfnamefont{T.}~\bibnamefont{Amand}},
  \bibinfo{author}{\bibfnamefont{L.}~\bibnamefont{Bouet}},
  \bibinfo{author}{\bibfnamefont{T.}~\bibnamefont{Kuroda}},
  \bibinfo{author}{\bibfnamefont{T.}~\bibnamefont{Mano}},
  \bibinfo{author}{\bibfnamefont{D.}~\bibnamefont{Paget}},
  \bibinfo{author}{\bibfnamefont{O.}~\bibnamefont{Krebs}},
  \bibinfo{author}{\bibfnamefont{X.}~\bibnamefont{Marie}},
  \bibinfo{author}{\bibfnamefont{K.}~\bibnamefont{Sakoda}},
  \bibnamefont{et~al.}, \bibinfo{journal}{Nat. Commun.}
  \textbf{\bibinfo{volume}{5}}, \bibinfo{pages}{3268} (\bibinfo{year}{2014}).

\bibitem[{\citenamefont{{Huo} et~al.}(2014)\citenamefont{{Huo}, {Witek},
  {Kumar}, {Cardenas}, {Zhang}, {Akopian}, {Singh}, {Zallo}, {Grifone},
  {Kriegner} et~al.}}]{2014NatPh..10...46H}
\bibinfo{author}{\bibfnamefont{Y.~H.} \bibnamefont{{Huo}}},
  \bibinfo{author}{\bibfnamefont{B.~J.} \bibnamefont{{Witek}}},
  \bibinfo{author}{\bibfnamefont{S.}~\bibnamefont{{Kumar}}},
  \bibinfo{author}{\bibfnamefont{J.~R.} \bibnamefont{{Cardenas}}},
  \bibinfo{author}{\bibfnamefont{J.~X.} \bibnamefont{{Zhang}}},
  \bibinfo{author}{\bibfnamefont{N.}~\bibnamefont{{Akopian}}},
  \bibinfo{author}{\bibfnamefont{R.}~\bibnamefont{{Singh}}},
  \bibinfo{author}{\bibfnamefont{E.}~\bibnamefont{{Zallo}}},
  \bibinfo{author}{\bibfnamefont{R.}~\bibnamefont{{Grifone}}},
  \bibinfo{author}{\bibfnamefont{D.}~\bibnamefont{{Kriegner}}},
  \bibnamefont{et~al.}, \bibinfo{journal}{Nat. Phys.}
  \textbf{\bibinfo{volume}{10}}, \bibinfo{pages}{46} (\bibinfo{year}{2014}).

\end{thebibliography}

\end{document}